\documentclass[sigconf]{acmart}
\AtBeginDocument{%
  }

\setcopyright{acmlicensed}

\acmISBN{978-1-4503-XXXX-X/2018/06}

\renewcommand\footnotetextcopyrightpermission[1]{}

\setcopyright{rightsretained}
\copyrightyear{2025}
\acmYear{2025}
\acmConference[GENNEXT@SIGIR'25]{}{July 17}{Padua, Italy}
\acmBooktitle{Proceedings of The 1st Workshop on Next Generation of IR and Recommender Systems with Language Agents, Generative Models, and Conversational AI (GENNEXT@SIGIR'25), July 17, 2025}
\acmYear{2025}
\acmMonth{07}
\acmDOI{}
\acmISBN{}
\acmPrice{15.00}

\begin{document}

\title{CAL-RAG: Retrieval-Augmented Multi-Agent Generation for Content-Aware Layout Design}


% ---------- AUTHORS ----------
\author{%
  Najmeh Forouzandehmehr,
  Reza Yousefi Maragheh,
  Sriram Kollipara,
  Kai Zhao,
  Topojoy Biswas,
  Evren Korpeoglu,
  Kannan Achan\\[4pt]
  Walmart Global Tech, Sunnyvale, California, USA\\[4pt]
  \texttt{\small
    \{najmeh.forouzandehmehr,
      reza.yousefimaragheh,
      sriram.kollipara,
      kai.zhao,
      topojoy.biswas,
      evren.korpeoglu,
      kannan.achan\}@walmart.com}}
% ---------- END AUTHORS ----------

\begin{comment}

\author{Najmeh Forouzandehmehr}
\affiliation{%
  \institution{Walmart Global Tech} 
  \city{Sunnyvale} 
  \state{CA} 
}
\email{najmeh.forouzandehmehr@walmart.com}

\author{Reza Yousefi Maragheh}
\affiliation{%
  \institution{Walmart Global Tech}
  \city{Sunnyvale} 
  \state{CA} 
  }
\email{reza.yousefimaraghehr@walmart.com}

\author{Sriram Kollipara}
\affiliation{%
  \institution{Walmart Global Tech}
  \city{Sunnyvale} 
  \state{CA} 
  }
\email{sriram.kollipara@walmart.com}

\author{Kai Zhao}
\affiliation{%
  \institution{Walmart Global Tech}
  \city{Sunnyvale} 
  \state{CA} 
  }
\email{kai.zhao@walmart.com}

\author{Topojoy Biswas}
\affiliation{%
  \institution{Walmart Global Tech}
  \city{Sunnyvale} 
  \state{CA} 
  }
\email{topojoy.biswas@walmart.com}

\author{Evren Korpeoglu}
\affiliation{%
  \institution{Walmart Global Tech}
  \city{Sunnyvale} 
  \state{CA} 
  }
\email{EKorpeoglu@walmart.com}

\author{Kannan Achan}
\affiliation{%
  \institution{Walmart Global Tech}
  \city{Sunnyvale} 
  \state{CA} 
  }
\email{kannan.achan@walmart.com}
\end{comment}
\renewcommand{\shortauthors}{Forouzandehmehr et al.}


\begin{abstract}
Automated content-aware layout generation—the task of arranging visual elements such as text, logos, and underlays on a background canvas—remains a fundamental yet underexplored problem in intelligent design systems. While recent advances in deep generative models and large language models (LLMs) have shown promise in structured content generation, most existing approaches lack grounding in contextual design exemplars and fall short in handling semantic alignment and visual coherence. In this work, we introduce CAL-RAG, a Retrieval-Augmented, Agentic framework for content-aware layout generation that integrates multimodal retrieval, large language models, and collaborative agentic reasoning. Our system retrieves relevant layout examples from a structured knowledge base and invokes an LLM-based layout recommender to propose structured element placements. A vision-language grader agent evaluates the layout based on visual metrics, and a feedback agent provides targeted refinements, enabling iterative improvement. We implement our framework using LangGraph and evaluate on the PKU PosterLayout dataset, a benchmark rich in semantic and structural variability. CAL-RAG achieves state-of-the-art performance across multiple layout metrics—including underlay effectiveness, element alignment, and overlap—substantially outperforming strong baselines such as LayoutPrompter. Our results demonstrate that combining retrieval augmentation with agentic multi-step reasoning provides a scalable, interpretable, and high-fidelity solution for automated layout generation.

\end{abstract}



\maketitle

\section{Introduction}

\begin{figure}[t]
  \centering
  \includegraphics[width=0.8\columnwidth]{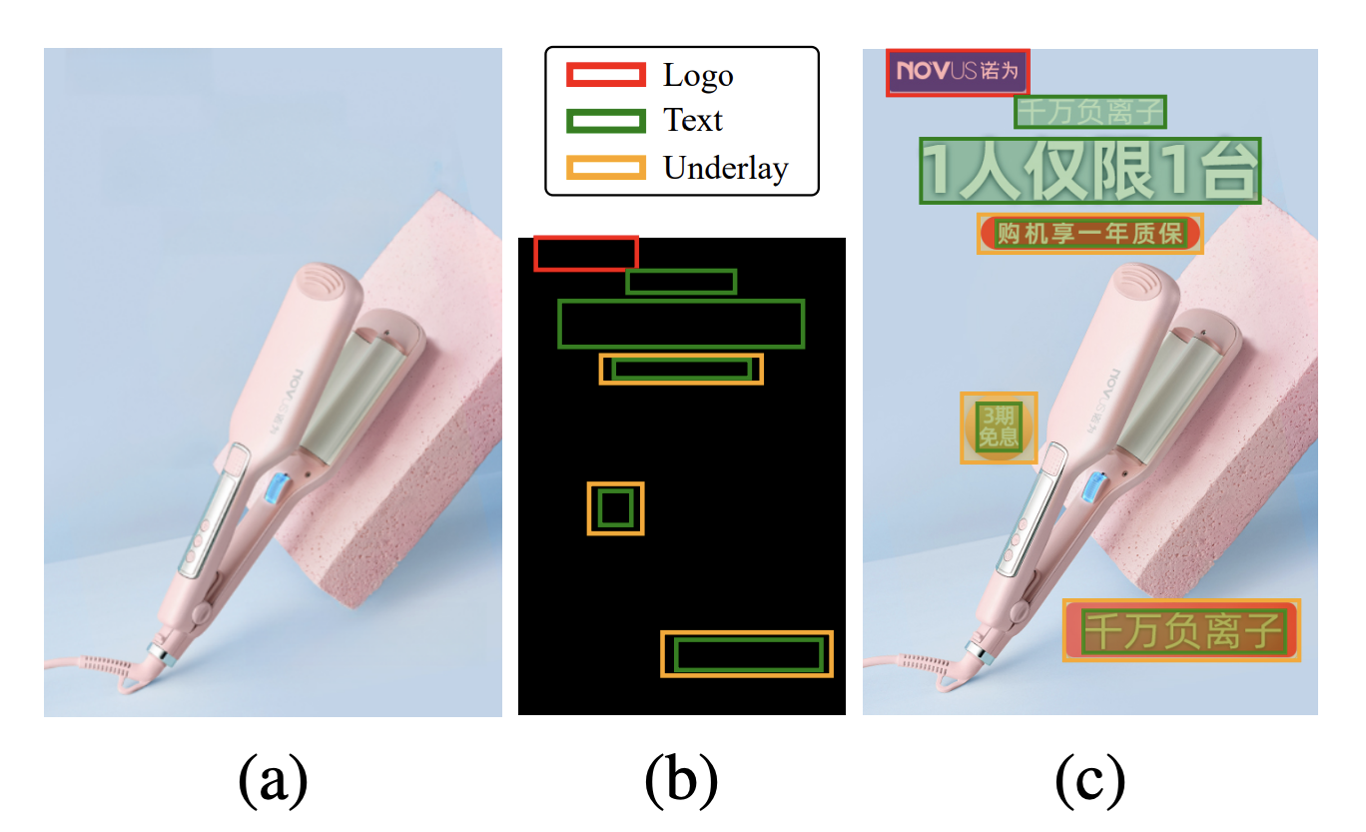}
  \caption{Example for content-aware layout generation: (a) Input canvas with background and content elements; (b) Generated layout based on visual and textual content awareness; (c) Final rendered presentation using the layout from (b).}
  \Description{}
  \label{fig:pku}
\end{figure}

\begin{figure*}[h]
  \centering
  \includegraphics[width=0.80\linewidth]{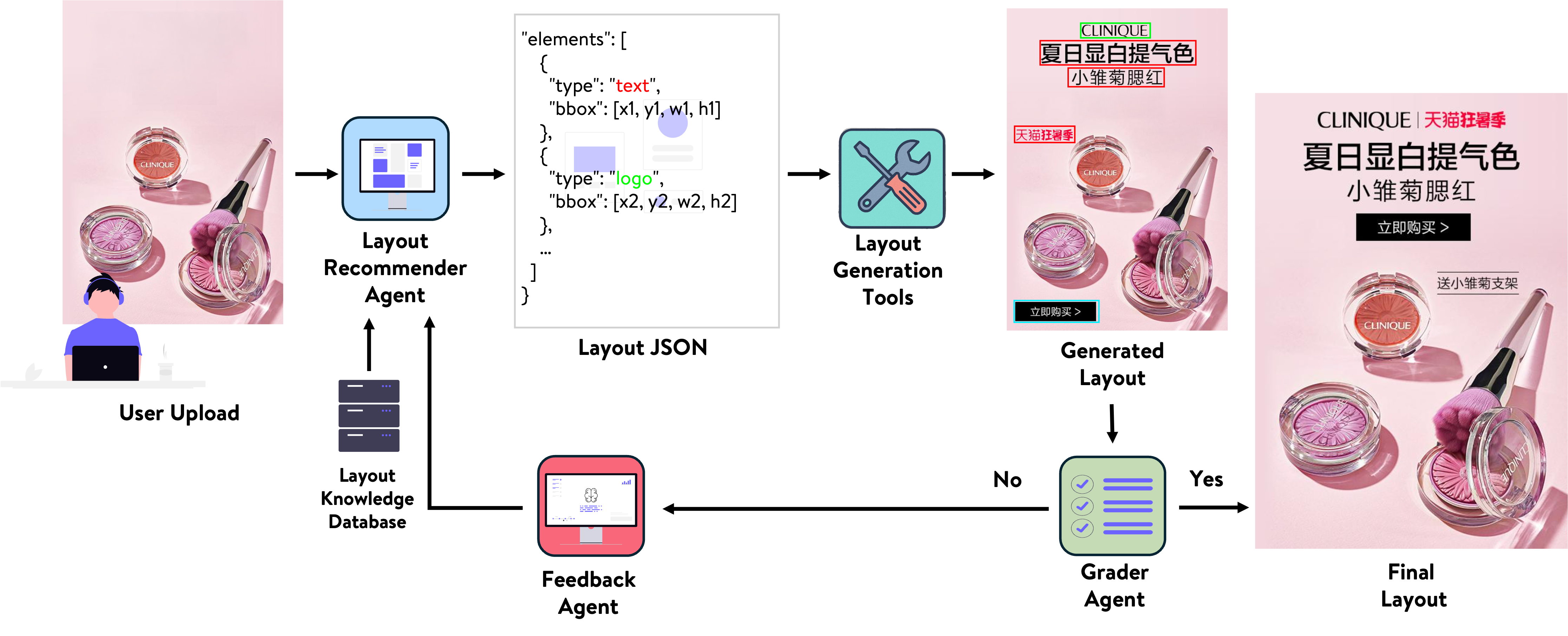}
  \caption{System Architecture Diagram of CAL-RAG.}
  \Description{}
  \label{fig:system_achitecture}
\end{figure*}

Designing content-aware visual layouts is a fundamental challenge in computational creativity, with applications spanning digital advertising, web interfaces, and e-commerce presentation. The task involves arranging heterogeneous visual elements—such as text, logos, and underlays—on a canvas in a way that balances aesthetic principles with semantic intent. Despite the practical importance of this task, existing methods often fall short in handling the combinatorial, content-sensitive nature of layout generation.

Traditional approaches either relied on rule-based heuristics or framed layout design as an optimization problem over hand-crafted alignment and spacing constraints \cite{zhang2024vascar}. These methods offer limited generalization, particularly in diverse or semantically rich design contexts. The advent of deep generative models—such as GANs, VAEs, and transformers—has enabled more expressive layout synthesis by learning from annotated layout corpora \cite{jyothi2019layoutvae, chai2023layoutdm, Li2019LayoutGAN,arroyo2021variational, horita2024retrieval}. However, these models are typically data-hungry, brittle to out-of-distribution inputs, and often struggle to incorporate visual-semantic alignment at inference time.

More recently, Large Language Models (LLMs) and Large Vision-Language Models (LVLMs) have demonstrated remarkable zero-shot capabilities in generating structured outputs, such as HTML or JSON representations of layouts, conditioned on multimodal inputs \cite{Feng2023LayoutGPT, lin2023layoutprompter, yang2024posterllava, Seol2024PosterLLAMA}. Yet, most LLM-driven methods rely on prompt engineering alone, lacking access to contextual design knowledge or external grounding signals. Retrieval-Augmented Generation (RAG) presents a promising solution by coupling generation with example-driven retrieval to enhance factuality and contextual alignment \cite{gao2023retrieval, horita2024retrieval, chen2024benchmarking}. A notable prior work in this direction is the Retrieval-Augmented Layout Transformer (RALF), which retrieves nearest neighbor layout examples based on an input image and feeds these results into an autoregressive generator, achieving state-of-the-art performance in content-aware layout generation \cite{horita2024retrieval}.

In this work, we propose CAL-RAG, a novel framework that integrates RAG with a collaborative, multi-agent system \cite{Gou2023CRITIC, Wu2023AutoGen,Singh2025Agentic, yan2024corrective} for content-aware layout generation. Our architecture is built upon three specialized agents: a layout recommender powered by an LLM, a vision-language grader agent, and a feedback agent that supports iterative refinement. The system is instantiated using LangGraph, and is trained and evaluated on the PKU PosterLayout dataset \cite{hsu2023posterlayout}, a challenging benchmark containing semantically diverse poster designs.

Unlike prior work, CAL-RAG grounds layout decisions in a retrieval corpus of design exemplars, enabling better inductive bias and interpretability. The layout generator reasons over retrieved samples to produce structured layout hypotheses; these are then rigorously scored by the grader agent using geometric and visual coherence metrics. If a layout is rejected, the feedback agent intervenes with localized corrections, prompting the recommender for an updated design. Through this agentic iteration loop, CAL-RAG achieves state-of-the-art performance across multiple layout quality metrics, including underlay effectiveness, alignment, and element separation. 
%To the best of our knowledge, this is the first work to jointly leverage retrieval augmentation, vision-language grading, and agentic feedback loops for automated layout generation. 
Our contributions are threefold:

\begin{itemize}
    \item We present a novel agentic RAG framework for layout generation that unifies retrieval, generation, evaluation, and revision within a single compositional system.
    \item We introduce a principled evaluation protocol based on geometric layout metrics and visual-semantic alignment on a challenging public benchmark.
    \item We demonstrate significant empirical gains over strong baselines, including LayoutPrompter and RALF, through extensive ablation and comparative analysis.
\end{itemize}

%Our results suggest that integrating multi-agent reasoning with retrieval and vision-language feedback enables scalable and high-fidelity layout design, paving the way for more controllable and interpretable creative AI systems.

The core problem we tackle is: Given a background canvas image, how can we generate a layout that specifies the number, type, and position of content elements (text, logo, underlay) in a way that balances semantic coherence, aesthetic principles, inspired by prior design patterns. The input to our system is a background image, from which the type and number of content elements (e.g., text, logos, underlays) are automatically inferred. The output is a structured layout proposal, defined as a set of bounding boxes with associated element types.

\if
The automated generation of content-aware layouts has emerged as a significant area of focus within graphic design intelligence, with implications for advertising, web design, and document creation\cite{Li2019LayoutGAN,Feng2023LayoutGPT}. A fundamental challenge lies in automatically producing layouts that are not only aesthetically pleasing but also functionally effective in conveying information while respecting the underlying visual content  \cite{Seol2024PosterLLAMA}. Traditionally, this task has been the domain of skilled human designers, requiring significant expertise and iterative refinement. The increasing demand for efficient and accessible design solutions has spurred research into automated methods that can democratize the design process and enhance productivity \cite{Seol2024PosterLLAMA,Tang2023LayoutNUWA}.

The automation of content-aware layout generation has evolved significantly over time. Early attempts often involved the application of optimization techniques. These methods were guided by a set of predefined design principles that dictated how visual elements should be arranged on a canvas. While these approaches could produce layouts adhering to specific rules, they often lacked the adaptability and creative flair of human designers, particularly when dealing with diverse content and complex semantic relationships.\cite{zhang2024vascar} 

The advent of deep learning brought about a paradigm shift in the field of layout generation. Researchers began to explore generative models capable of learning the underlying distribution of layouts from training data. This led to the development of various architectures, including Generative Adversarial Networks (GANs), Variational Autoencoders (VAEs), Transformer-based models, and diffusion-based models \cite{jyothi2019layoutvae, arroyo2021variational, horita2024retrieval, chai2023layoutdm}. These models aimed to synthesize novel layouts that not only looked aesthetically pleasing but also effectively conveyed information based on the input content.   

For instance, the Composition-aware Graphic Layout GAN (CGL-GAN) was developed to utilize the visual content of input images to generate corresponding layouts \cite{zhou2022composition}. This model demonstrated the importance of considering the semantic information present in the visual content during the layout generation process. Similarly, the Design Sequence GAN (DS-GAN) introduced the concept of design sequences to guide the generation process, attempting to mimic the step-by-step approach that human designers often take \cite{hsu2023posterlayout}. Transformer-based models, such as LayoutDM, leveraged self-attention mechanisms to capture the contextual relationships between different layout elements, allowing for a more holistic understanding of the layout structure\cite{chai2023layoutdm}.   

Recent research has highlighted the surprising ability of Large Language Models (LLMs) to generate structured outputs, such as HTML or JSON, which can represent visual layouts \cite{Seol2024PosterLLAMA, Feng2023LayoutGPT, lin2023layoutprompter}. This capability suggests that LLMs can potentially solve layout generation as a programming task, likely learned from design-related code samples within their training corpora [2, 4]. Pioneering works like LayoutPrompter \cite{lin2023layoutprompter} and LayoutGPT \cite{Feng2023LayoutGPT} have demonstrated the feasibility of using LLMs for layout generation through in-context learning, showcasing their adaptability to this domain without explicit visual training. Furthermore, the evolution of LLMs into Large Vision-Language Models (LVLMs) with enhanced multi-modal understanding capabilities presents promising avenues for advancing content-aware layout generation \cite{yang2024posterllava}.

To further enhance the performance of generative models, including those used for creative tasks, the technique of Retrieval-Augmented Generation (RAG) has gained significant prominence \cite{gao2023retrieval, chen2024benchmarking}. RAG combines the strengths of retrieval-based methods with generative models by first fetching relevant external information, such as documents or data snippets, based on a given query. This retrieved information is then used to augment the input provided to the LLM, enabling it to generate more accurate and contextually relevant outputs. By grounding their responses in external knowledge, LLMs can mitigate issues like outdated information and factual inaccuracies.   

The potential of RAG extends to creative tasks by providing generative models with relevant examples and inspiration. In domains like layout generation, where obtaining large, high-quality annotated datasets can be challenging, the principle of RAG can be adapted to retrieve relevant layout examples based on the input content . This approach can provide the generative model with useful references and potentially address the data scarcity problem. A notable prior work in this direction is the Retrieval-Augmented Layout Transformer (RALF), which retrieves nearest neighbor layout examples based on an input image and feeds these results into an autoregressive generator, achieving state-of-the-art performance in content-aware layout generation \cite{horita2024retrieval}.

Agentic frameworks in AI represent a paradigm where AI agents can autonomously determine and execute a sequence of actions to achieve a specific goal. Key characteristics of these agents include memory, planning capabilities, and the ability to utilize tools to interact with their environment \cite{Gou2023CRITIC, Wu2023AutoGen,Singh2025Agentic, yan2024corrective},. The concept of agentic RAG has recently emerged, integrating AI agents into the RAG pipeline to enhance its adaptability and accuracy \cite{Singh2025Agentic,yan2024corrective}. This integration allows for more sophisticated retrieval strategies, multi-step reasoning, and the ability to leverage multiple information sources. While the use of LLMs and RAG for layout generation has been explored, this research distinguishes itself by introducing a collaborative agentic framework where a layout generator, a visual grader, and a feedback provider work in concert to produce high-quality, content-aware layouts.
This research introduces a novel approach that integrates a RAG-based framework with an agentic model for the task of content-aware layout generation. This combination aims to leverage the benefits of both techniques: RAG's ability to access and utilize relevant layout examples from a knowledge base, and the agentic framework's capacity for autonomous decision-making and task execution . The system is implemented using Langgraph, a framework designed for building LLM-powered applications. The model utilizes a RAG database constructed from the PKU PosterLayout dataset, which is specifically designed for content-aware visual-textual presentation layout and defines element types such as text, logo, and underlay. An LLM agent acts as a layout recommender, guided by a carefully crafted prompt that instructs it to analyze retrieved examples and the input canvas to suggest the number and types of elements, along with their corresponding bounding boxes, in a structured JSON format. Crucially, the agentic framework also incorporates a VLM grader agent responsible for evaluating the generated layout against a set of visual metrics criteria. If the grader deems the layout unacceptable, a VLM feedback agent steps in to analyze the layout and provide specific feedback to the initial layout generator, enabling an iterative refinement process. The performance of the generated layouts is evaluated using metrics such as underlay effectiveness (loose and strict), overlay, and alignment. This work contributes to the field by exploring the novel application of a collaborative multi-agent RAG framework to content-aware layout generation, potentially leading to improved layout quality, enhanced content awareness, and increased data efficiency. The subsequent sections of this paper will delve into the architecture and implementation of the proposed model, the experimental setup, the results obtained, and a discussion of the findings and their implications.

\fi

\section{Methodology}
Our approach to content-aware layout generation (CAL-RAG) leverages a multi-agent system integrated with a Retrieval-Augmented Generation (RAG) framework. The overall system architecture comprises four key components: a Layout Recommender Agent, a Layout Generation Tool, a Grader Agent, and a Feedback Agent (see Figure. \ref{fig:system_achitecture}). This framework facilitates an iterative process of layout creation and refinement, aiming to produce high-quality, content-aware visual-textual presentations.

\subsection{Layout Recommender Agent}
\begin{comment}

The process begins with the \textbf{Layout Recommender Agent}, which is responsible for suggesting an initial layout for a given background image. This agent utilizes the \textit{PKU PosterLayout} dataset as its layout knowledge base. Given a background image, the agent first retrieves a set of semantically similar layouts from the knowledge base. The retrieval mechanism employs a technique such as \textit{embedding similarity} to identify layouts with visual characteristics comparable to the input background.

Drawing inspiration from these retrieved examples and guided by established design principles, the Layout Recommender Agent determines an appropriate number and type of visual elements (e.g., text, logo, underlay) for the canvas. It then predicts bounding boxes for each element, ensuring visual coherence, balance, proper spacing, and alignment, consistent with the examples retrieved.

To guide this decision-making process, the agent is given structured instructions that define how to:
\begin{itemize}
    \item Analyze retrieved layout examples alongside the test canvas,
    \item Suggest the number and types of elements needed,
    \item Assign bounding boxes for these elements, ensuring spatial harmony and alignment.
\end{itemize}

The agent's output is a JSON object that includes the canvas dimensions and a list of elements, each specified by its type and bounding box coordinates.}
\end{comment}

The process begins with the \textbf{Layout Recommender Agent}, which suggests an initial layout $L$ for a given background image $I_{\text{bg}}$. This agent uses the \textit{PKU PosterLayout} dataset as its layout knowledge base. This dataset consists of pairs of background image and ground truth layouts. Let us denote the set of these pairs by:  $\mathcal{D} = \{(I_{j}, L_{j}), \textit{for } j \in \mathcal{N}\}$, when $\mathcal{N}$ is set of indexes. 
After retrieving k similar examples from $\mathcal{D}$, the layout recommender uses them as few-shot references to guide an unconstrained generation process. Given these examples, it predicts the number of visual elements (e.g., logos, text blocks, overlays), their types, and the corresponding bounding boxes—producing a layout tailored to the input canvas.
An example of the unconstrained layout generation task using the PKU dataset is shown in Figure~\ref{fig:pku}.
\begin{comment}

At a high level, the layout recommender agent retrieves a set of similar layouts from $\mathcal{D}$ and then predicts bounding boxes for visual elements (e.g., text blocks, logos, underlays) that are spatially coherent with the input canvas.
\end{comment}
\paragraph{1. Embedding and Retrieval.}

We assume each background image $I$ can be transformed into a latent vector (embedding) via an embedding function: $\mathbf{E}(I) \in \mathbb{R}^d$. We use the CLIP image encoder for for $\mathbf{E}(.)$ \cite{radford2021learning}.
\begin{comment}

We assume each background image $I$ can be transformed into a latent vector (embedding) via an embedding function: $\mathbf{E}(I) \in \mathbb{R}^d.$ We use {\color{blue} clip embbedding function??} for $\mathbf{E}(.)$. {\color{blue}CITE}
\end{comment}
Given an input background image $I_{\text{bg}}$, the agent computes its embedding $\mathbf{E}(I_{\text{bg}})$. For each candidate layout-image pair $\bigl(I_i, L_i\bigr)$ in the knowledge base, it also has a corresponding image embedding $\mathbf{E}(I_i)$. The retrieval step is conducted by a cosine similarity score $S$ between the background image and a reference image in the dataset:
\[
S\bigl(I_{\text{bg}}, I_i\bigr) 
= \frac{\mathbf{E}(I_{\text{bg}}) \cdot \mathbf{E}(I_i)}{\lVert \mathbf{E}(I_{\text{bg}})\rVert \,\lVert \mathbf{E}(I_i)\rVert}.
\]
{ CAL-RAG} then selects set the top-$k$ most similar reference layouts:
\begin{equation*}
    \mathcal{R} = \Bigl\{ (I_i, L_i) ; 
i \in \mathrm{argtop}_k \{S(I_{\text{bg}}, I_m)\}_{m \in \mathcal{N}} 
\Bigr\}.
\end{equation*}
Set $\mathcal{R}$ acts as source for ``best practices'' of bounding-box positioning, number of elements, and arrangement for the layout instance at hand.

\paragraph{2. Bounding Box Estimation.}
Let the agent decide on $n$ elements to be placed in the poster layout, each denoted as $e_j$ for $j=1,\ldots,n$. Each element $e_j$ (e.g., text, logo, underlay) is represented by a bounding box:
\[
B_j = \bigl(x_j, \, y_j, \, w_j, \, h_j\bigr),
\]
where $(x_j, \, y_j)$ is the top-left coordinate of the bounding box and $(w_j, \, h_j)$ are the width and height, respectively.

The placement of these $n$ bounding boxes constitutes a layout $L = \{ B_1, B_2, \ldots, B_n \}$. To incorporate design principles such as balance and spatial harmony, the agent follows structured instructions and uses features from retrieved layouts $\mathcal{R}$. For example, it might minimize an overall \emph{layout cost} $\mathcal{C}(L)$ that accounts for alignment, overlap, and aesthetic metrics:
\[
\mathcal{C}(L) 
= \alpha_1 \mathcal{C}_{\text{overlap}}(L) 
+ \alpha_2 \mathcal{C}_{\text{alignment}}(L)
+ \alpha_3 \mathcal{C}_{\text{margins}}(L),
\]
where $\mathcal{C}_{\text{overlap}}(L)$ measures the total unwanted overlap among bounding boxes, $\mathcal{C}_{\text{alignment}}(L)$ measures alignment deviations among box edges (e.g., difference in $x$-coordinates for left alignment), $\mathcal{C}_{\text{margins}}(L)$ captures violations of spacing constraints from the design guidelines, $\alpha_1, \alpha_2, \alpha_3$ are weighting hyperparameters reflecting the importance of each principle.

Hence, the Layout Recommender Agent seeks:
\[
L^* \;=\; \underset{L}{\mathrm{argmin}} \;\mathcal{C}(L),
\]
subject to any constraints derived from the retrieved examples (e.g., recommended number of elements, recommended bounding box aspect ratios).

\paragraph{3. Structured Instructions and Output.}
To ensure reproducibility and clarity, the agent is given \emph{structured instructions} that define how to: \textbf{Analyze} the retrieved layouts $\mathcal{R}$ and compare them with the test canvas $I_{\text{bg}}$; \textbf{Determine} the number of elements $n$ and their types (e.g., text, logo); \textbf{Assign} bounding boxes $B_1, \dots, B_n$ while adhering to spatial alignment and design heuristics.

\if0
After these steps, the agent's output is a JSON object of the following form:

\begin{verbatim}
{
  "canvas": {
    "width": W,
    "height": H
  },
  "elements": [
    {
      "type": "text",
      "bbox": [x1, y1, w1, h1]
    },
    {
      "type": "logo",
      "bbox": [x2, y2, w2, h2]
    },
    ...
  ]
}
\end{verbatim}

Here, $(W, H)$ are the dimensions of the input canvas, and each array $[x_j, y_j, w_j, h_j]$ corresponds to the bounding box $B_j$. By weaving together the retrieved insights $\mathcal{R}$, design principles, and a cost-minimization procedure, the Layout Recommender Agent efficiently proposes a layout $L^*$ that ensures aesthetic consistency, visual balance, and content relevance.
\fi

\subsection{Layout Generation Tool}

%The JSON output from the Layout Recommender Agent is then passed to a dedicated Layout Generation Tool. This module is responsible for physically realizing the suggested layout by placing the specified elements (text, logo, underlay) onto the provided background image according to the recommended bounding box coordinates. This step transforms the abstract layout recommendations into a concrete visual representation that can be further evaluated. 

After the \textbf{Layout Recommender Agent} produces its JSON output, the recommended bounding boxes $\{B_1, \ldots, B_n\}$ and associated element types (e.g., text, logo, underlay) are passed to the Layout Generation Tool. Let the background image be denoted by $I_{\text{bg}}$. The goal of this tool is to physically realize the abstract layout $L = \{ B_j : j=1,\dots,n \}$ by compositing the specified elements onto $I_{\text{bg}}$ according to their bounding box coordinates.

\paragraph{1. Compositing Framework.}

We model each visual element as a function or image $C_j$ (e.g., text rendered as a small image or graphic, a logo image, etc.). Thus, the Layout Generation Tool aims to produce a final composed image $I_{\text{comp}}$ by placing each $C_j$ into the bounding box $B_j = (x_j, \, y_j, \, w_j, \, h_j)$ on the canvas of size $(W, H)$. Formally, we define a compositing function
\[
\mathcal{F}\bigl(I_{\text{bg}}, \{(C_j, B_j)\}_{j=1}^{n}\bigr) \;=\; I_{\text{comp}},
\]
where $\mathcal{F}$ positions each element $C_j$ in $B_j$ and at location $(x_j, y_j)$, scales it if necessary to fit width $w_j$ and height $h_j$, and overlays or blends it (depending on the element type) onto $I_{\text{bg}}$.

\paragraph{2. Coordinate Transform for Placing Elements.}

For each element $C_j$, which may have an intrinsic resolution or size, we define a coordinate transform $\Phi_j$ that maps the local coordinates of $C_j$ to the global coordinates of $I_{\text{bg}}$ according to $B_j$. If $C_j$ has a native size $(w_j^0,\, h_j^0)$, then placing it into the bounding box $B_j=(x_j,y_j,w_j,h_j)$ involves a scaling factor $(w_j/w_j^0, h_j/h_j^0)$. 

\paragraph{3. Overlay and Blending Operations.}

To combine $C_j$ with the background, we can use a simple overwrite or alpha-blending operation. For example, a standard ``over'' operator from alpha compositing can be defined for each pixel $\mathbf{q}$ in $I_{\text{bg}}$:
\[
I_{\text{comp}}(\mathbf{q})
=\; (1 - \alpha_j(\mathbf{q})) \, I_{\text{bg}}(\mathbf{q})
\;+\; \alpha_j(\mathbf{q}) \, C_j\bigl(\Phi_j^{-1}(\mathbf{q})\bigr),
\]
where $\alpha_j(\mathbf{q}) \in [0,1]$ is the opacity of element $C_j$ at pixel $\mathbf{q}$. If $C_j$ contains no transparency, $\alpha_j(\mathbf{q}) = 1$ for all valid $\mathbf{q}$ in the bounding box, effectively overwriting the background in that region.

\paragraph{4. Concrete Implementation.}

For $n$ elements, the compositing function $\mathcal{F}$ applies the above placement and blending steps in a chosen order (e.g., from back to front). Hence, the final composition can be written as a recursive or sequential update:
\[
I_{\text{comp}}^{(j)} \;=\; \mathrm{Blend}\Bigl(I_{\text{comp}}^{(j-1)},\, C_j,\, B_j\Bigr),
\]

The Layout Generation Tool transforms the abstract layout recommendations into a concrete visual representation that can be subsequently evaluated or refined by other agents in the system.

\subsection{Grader Agent}

{Once the layout $L$ is generated by the Layout Generation Tool, it is evaluated by the \textbf{Grader Agent}, which acts as a graphic design expert. This agent measures the quality and acceptability of $L$ based on a set of predefined visual metrics. Let these metrics be represented by the function
\[
\Gamma(L) \;=\; \bigl(\gamma_{1}(L),\, \gamma_{2}(L),\, \gamma_{3}(L),\, \gamma_{4}(L)\bigr),
\]
where each component $\gamma_{k}(L)$ corresponds to a specific criterion:

 \textbf{Consistency and Cohesion} ($\gamma_{1}(L)$): 
    \[
    \gamma_{1}(L) \;=\; \exp\bigl(-\sigma_{\mathrm{colors}}(L)\bigr),
    \]
    where $\sigma_{\mathrm{colors}}(L)$ measures the standard deviation of dominant colors across all placed elements (lower deviation implies higher cohesion). Hence, $\gamma_{1}(L) \in (0,1]$ increases as color usage becomes more harmonious.

     \textbf{Composition and Spacing} ($\gamma_{2}(L)$):  
    \[
    \gamma_{2}(L) \;=\; 1 - \frac{\text{OverlapArea}(L)}{\text{TotalElementArea}(L)},
    \]
    where $\text{OverlapArea}(L)$ sums the pairwise overlapping regions of bounding boxes, and $\text{TotalElementArea}(L)$ is the combined area of all bounding boxes (without subtracting overlaps). A well-spaced layout has minimal overlap, making $\gamma_{2}(L)$ closer to 1.

     \textbf{Product Visibility and Clarity} ($\gamma_{3}(L)$):  
    \[
    \gamma_{3}(L) 
    \;=\; 1 - \frac{\sum_{j=1}^{n} \mathbb{I}\{\mathrm{Occluded}(B_j)\}}{n},
    \]
    where $\mathbb{I}\{\mathrm{Occluded}(B_j)\}$ is an indicator function that is 1 if bounding box $B_j$ is significantly occluded (e.g., by another element) and 0 otherwise, and $n$ is the total number of elements. Higher clarity yields fewer occlusions, pushing $\gamma_{3}(L)$ toward 1.
  
Let $\mathbf{t} = (t_{1}, t_{2}, t_{3})$ be threshold values for these metrics. The Grader Agent outputs:
\[
\text{Decision}(L) \;=\;
\begin{cases}
\texttt{Accept}, & \text{if } \gamma_{k}(L) \,\ge\, t_{k} \;\; \forall\,k=1,2,3, \\[6pt]
\texttt{Reject}, & \text{otherwise}.
\end{cases}
\]
Thus, the layout is \texttt{Accept}ed only if it meets or exceeds \emph{all} metric thresholds; otherwise, the Grader Agent outputs \texttt{Reject}.}

\subsection{Feedback Agent}

If the \textbf{Grader Agent} rejects the generated layout $L$, the system invokes the \textbf{Feedback Agent} for targeted improvements. Acting as a graphic design expert, this agent analyzes the rejected layout to identify specific areas for refinement. In particular, it focuses on:
\begin{itemize}
    \item \textbf{Placement} and \textbf{alignment} of elements for balanced composition,
    \item \textbf{Cohesiveness} and \textbf{proportions} of bounding boxes,
    \item Maintaining designated empty space in a certain canvas region $\Omega \subset \text{Canvas}$,
    \item Avoiding comments on \emph{color}, \emph{background}, or \emph{item selection}.
\end{itemize}

Formally, let $\Phi(L)$ denote the \emph{feedback function}, which encodes the agent's critique of $L$ and prescribes modifications. That is,
\[
\Phi(L) \;=\; \bigl(\delta_1(L),\, \delta_2(L),\, \ldots,\, \delta_n(L)\bigr),
\]
where each $\delta_j(L)$ provides a corrective shift (e.g., in position, size, or alignment) for the $j$-th bounding box of $L$. Adhering to the requirement that $\Omega$ remains clear, any suggested change for $B_j \in L$ must satisfy the constraint $B_j \cap \Omega = \varnothing$, if $B_j$ should not intrude on the empty space.

Once $\Phi(L)$ is computed, the layout is refined by the \textbf{Layout Recommender Agent}:
\[
L^{(t+1)} \;=\; \text{Refine}\!\Bigl(L^{(t)},\, \Phi\bigl(L^{(t)}\bigr)\Bigr),
\]
where $L^{(t)}$ is the layout at iteration $t$ and $\Phi\bigl(L^{(t)}\bigr)$ contains the feedback from the \emph{Feedback Agent} regarding placement and alignment. This iterative process repeats until the \textbf{Grader Agent} finally accepts the layout or a maximum number of iterations is reached.

\section{Experiments}

\subsection{Dataset and Evaluation Metrics}

The RAG database and the evaluation of the proposed model are based on the PKU PosterLayout dataset, a new benchmark specifically designed for content-aware visual-textual presentation layout \cite{hsu2023posterlayout}. This dataset comprises 9,974 poster-layout pairs and 905 image canvases with annotations defining three primary element types: text, logo, and underlay User Query. 
%The PKU dataset is designed to be challenging, exhibiting domain diversity, content diversity across nine product categories, and a high degree of layout variety and complexity, including layouts with more than 10 elements . The data was collected from various sources, including e-commerce poster datasets and image bank websites .
Following the experimental framework of RALF, the dataset is split into training (7,735), validation (1,000), and testing (1,000) sets . For easier utilization, the dataset provides original posters, inpainted posters (where visual elements are removed), image canvases, and saliency maps. 

The performance of the proposed model is evaluated using several layout quality metrics, including \textit{underlay effectiveness} (both loose and strict), \textit{overlay}, and \textit{alignment} \cite{lin2023layoutprompter}.

\begin{itemize}
    \item \textbf{Underlay Effectiveness} measures how well the generated underlay elements support and decorate non-underlay content, such as text or logos. A predicted underlay is considered valid if it effectively overlaps with at least one non-underlay element. The loose variant ($\text{Und}_{\ell}$)  rewards underlays that significantly intersect with non-underlay elements, taking the highest degree of overlap as the score. The strict variant ($\text{Und}_{s}$) is more conservative: it only gives credit if a non-underlay element is entirely contained within an underlay. $\text{Und}_{s}$ is then computed as the average score across all underlays.

    \item \textbf{Overlay} measures the average area of undesired overlap between non-underlay elements (e.g., text blocks, logos). It is computed as the ratio of the total overlapping area to the total layout area; lower values indicate better element separation and visual clarity.

    \item \textbf{Alignment} assesses how well elements are spatially aligned along common axes (e.g., left/right edges, centers). Misalignment is calculated as the average deviation between corresponding edge coordinates of adjacent elements. Lower scores indicate stronger geometric regularity and design coherence.
\end{itemize}
    

\subsection{Overall Results}

We compare the following methods in the experiments. CGL-GAN \cite{zhou2022composition} is a non-autoregressive encoder–decoder model employing a Transformer architecture. The model takes in an empty layout or layout constraints to decode a layout representation. 
DS-GAN [18] is a non-autoregressive model that utilizes a CNN-LSTM architecture \cite{hsu2023posterlayout}
LayoutPrompter, is a training-free method that leverages LLM prompting and in-context learning to tackle various layout generation tasks \cite{lin2023layoutprompter}.

\begin{table}[h]
    \centering
    \caption{Performance Comparison}
    \label{tab:geometric_comparison}
    \begin{tabular}{lcccc}
        \toprule
        \textbf{Methods} & Ove $\downarrow$ & Ali $\downarrow$ & Und$_l$ $\uparrow$ & Und$_s$ $\uparrow$ \\
        \midrule
        \textbf{Ground-Truth} & 0.0001 & 0.0002 & 0.9965 & 0.9912 \\
        \midrule
        \multicolumn{5}{l}{\textbf{Geometric Methods}} \\
        CGL-GAN & 0.0605 & 0.0062 & 0.8624 & 0.4043 \\
        DS-GAN  & 0.0220 & 0.0046 & 0.8315 & 0.4320 \\
        LayoutPrompter & 0.0036 & 0.0036 & 0.8986 & 0.8802 \\
         \textbf{CAL-RAG(Ours)} & \textbf{0.0023} & \textbf{0.002} & \textbf{1.0000} & \textbf{1.0000} \\
        \bottomrule
    \end{tabular}
\end{table}

%We evaluate CAL-RAG against a suite of strong baselines using four standard geometric layout metrics: Overlap (Ove), Alignment (Ali), and Underlay Effectiveness for large (Und<sub>l</sub>) and small (Und<sub>s</sub>) elements. Lower values of Ove and Ali indicate better spatial separation and alignment, respectively, while higher Und<sub>l</sub> and Und<sub>s</sub> scores reflect improved semantic positioning of underlay components.

As shown in Table \ref{tab:geometric_comparison}, CAL-RAG achieves state-of-the-art performance across all metrics, significantly outperforming LayoutPrompter and previous generation-based methods such as CGL-GAN and DS-GAN. We conducted evaluations following the PosterLayout setting  \cite{hsu2023posterlayout} and included previously reported results for comparison. Notably, CAL-RAG attains perfect underlay effectiveness (1.0000) for both large and small elements, indicating a strong alignment between generated layouts and the intended spatial semantics of content elements. Moreover, our method achieves the lowest overlap (0.0023) and alignment (0.002) among all models, closely approaching the ground-truth layout statistics, demonstrating the advantage of incorporating agentic iteration and retrieval-based grounding. These results validate the effectiveness of our retrieval-augmented, multi-agent design in generating layouts that are not only visually coherent but also semantically structured, setting a new benchmark for content-aware layout generation.

%As demonstrated in Table \ref{tab:geometric_comparison}, our agentic RAG model outperforms other baseline methods on key geometric metrics. Specifically, the Ove (Overlap) and Ali (Alignment) metrics show significant improvements compared to methods like LayoutPrompter and DS-GAN, with our model achieving the lowest values for both. Moreover, our model excels in Und$_l$ (underlap for large elements) and Und$_s$ (underlap for small elements), attaining perfect scores of 1.0000 in both categories, highlighting its superior ability to generate layouts with optimal spatial coherence and minimal underlap.

\subsection{Ablation Study}

To isolate the contributions of each component in the CAL-RAG architecture, we perform an ablation study by evaluating three progressively enhanced configurations: (1) Layout Recommender only, (2) Layout Recommender + Grader Agent, and (3) the full system with Grader and Feedback Agents. As shown in Table \ref{tab:ablation}, the base configuration, Layout Recommender Agent achieves modest performance. This suggests that while retrieval-based generation provides a useful prior, it lacks the self-corrective mechanisms necessary for fine-grained layout optimization.

\begin{table}[t]
  \centering
  \caption{Ablation Results}
  \label{tab:ablation}
  \begin{tabular}{lcccc}
    \toprule
    \textbf{Configuration} & Ove $\downarrow$ & Ali $\downarrow$ & Und$_l$ $\uparrow$ & Und$_s$ $\uparrow$ \\
    \midrule
    Layout Recommender only & 0.00341 & 0.0031 & 0.891 & 0.8 \\
    + Grader Agent          & 0.0029 & 0.0029 & 0.98 & 0.967 \\
    + Feedback Agent        & \textbf{0.0023} & \textbf{0.002} & \textbf{1}  & \textbf{1}\\
    \bottomrule
  \end{tabular}
\end{table}

Introducing the Grader Agent significantly boosts performance, particularly in underlay effectiveness, confirming that the grader effectively filters suboptimal layouts based on learned visual metrics. When the Feedback Agent is incorporated, the system achieves state-of-the-art scores across all metrics, including perfect alignment with semantic expectations (1.0000) and minimal overlap (0.0023). This demonstrates the crucial role of agentic feedback in iteratively refining layouts, validating our core hypothesis that a multi-agent loop of generation, evaluation, and correction yields superior results over single-pass generation methods.

\bibliographystyle{ACM-Reference-Format}
\bibliography{sample-base}


\end{document}
\endinput
%%
%% End of file `sample-sigconf-authordraft.tex'.